\documentclass[aps,twocolumn,twosides]{revtex4}%
\usepackage{graphicx,amssymb,amsmath,color,psfrag}
\usepackage{amsthm}
\usepackage{amsfonts}
\usepackage{algorithmic}
\usepackage{enumerate}
\usepackage{latexsym}
\usepackage{amsmath}
\usepackage{amssymb}
\usepackage{textcomp}
\usepackage{graphicx}
\usepackage{textcomp}
\usepackage{soul}
\usepackage{color}
\providecommand{\U}[1]{\protect\rule{.1in}{.1in}}

\begin{document}

\title{Black hole optical analogue: photon sphere microlasers}

\author{Chenni Xu$^{1\mathsection}$}
\author{Aswathy Sundaresan$^{1\mathsection}$}
\author{Nazire-Beg\"um Kazkal$^{2}$}
\author{Clement Lafargue$^{3}$}
\author{Lior Zarfaty$^{1}$}
\author{Li-Gang Wang$^{4}$}
\author{Ofek Birnholtz$^{1}$}
\author{Dominique Decanini$^{2}$}
\author{Melanie Lebental$^{2}$}
\author{Patrick Sebbah$^{1}$}
\thanks{patrick.sebbah@biu.ac.il}
\affiliation{$^{1}$ Department of Physics, The Jack and Pearl Resnick Institute for Advanced Technology, Bar-Ilan University, Ramat-Gan 5290002, Israel}
\affiliation{$^{2}$Université Paris-Saclay, CNRS, Centre de Nanosciences et de Nanotechnologies, 91120, Palaiseau, France}
\affiliation{$^{3}$ Laboratoire Lumi\`{e}re, Mati\`{e}re et Interfaces (LuMIn) CNRS, ENS Paris-Saclay, Universit\'{e} Paris-Saclay,
CentraleSup\'{e}lec, 91190 Gif-sur-Yvette, France}
\affiliation{$^{4}$ School of Physics, Zhejiang University, Hangzhou 310058, China}

\begin{abstract}
The bell-like ringdown of the gravitational field in the last stage of the merging of massive black holes is now routinely detected on earth by the last generation of gravitational wave detectors.
Its spectrum is interpreted as a sum of damped sinusoidal vibrations of the spacetime in the vicinity of the black hole. These so-called quasinormal modes are currently the subject of extensive studies, yet, their true nature remains elusive.
Here, we emulate, in the laboratory, genuine four-dimension black hole metrics by two-dimensional optical curved surfaces that preserve the features of lightlike geodesics.
We analytically compute the optical quasinormal modes and show that they are confined around the photon sphere, the unstable region around a black hole where spacetime curvature traps light in circular orbits.
By 3D-printing non-Euclidean dye-doped microcavities, we demonstrate lasing at the photon sphere with a mode profile that closely matches the analytical prediction. These results paves the way for observing astrophysical phenomena in tabletop setups and is likely to inspire innovative designs in photonics.

\end{abstract}
\maketitle
\def\thefootnote{$\mathsection$}\footnotetext{These authors contributed equally to this work.}




\section{Introduction}
The merger of binary black holes (BHs) releases a tremendous amount of energy into the universe as gravitational waves, first detected in 2015 by the LIGO/VIRGO collaboration \cite{LIGO, GWTC-3, GWTC-3-TGR}. Their spectral signature has provided the opportunity to test general relativity and to indirectly estimate the mass and angular momentum of the BH \cite{Krishnan23}.
The complex frequencies that compose the ringdown spectrum observed at the last stage of BH merger, commonly described by BH perturbation theory \cite{review, Guo2022}, would correspond to quasinormal modes intuitively interpreted as damped oscillations of the spacetime, confined around the photon sphere (PS) of the BH \cite{review, YBChen, Schutz1985}.
The PS is a spherical region outside the BH, where spacetime is so extremely warped that light trajectories bend and get trapped in a circular orbit around the BH. Orbiting light rays with impact parameters slightly departing from the photon capture radius \cite{EHT}, will either get bent away from or engulfed into the BH.
The instability of this classical circular orbit raises a fundamental question: If this region is unstable, why should a gravitational mode be localized, or "scarred," around it? This challenge calls for a more precise characterization of these damped modes, including their spatial distribution.


In a controlled laboratory environment analog models of general relativity could serve as a platform to track down PS and light dynamics thereabouts. For instance, artificial optical materials have been used to achieve light bending \cite{ShengNP}, wavefront shaping \cite{ShengNC} and optical trapping \cite{elight}, with the index profiles derived from carefully engineered metrics \cite{Genov2009}.
Another table-top analog model of gravitational fields consists of a two-dimensional (2D) curved surfaces embedded in three-dimensional (3D) space \cite{Batz2008, Batz2010}, which are derived by considering a fixed time coordinate in the four-dimensional (4D) curved spacetime.
Based on this static projection, various optical phenomena have been investigated both theoretically and experimentally
\cite{Bekenstein2014, Ulf2015, Bekenstein2017, Xu2018, Bekenstein2018, Xu2021, Xu2023, Ding2023, Zhang2024}. However, removing the time coordinate by setting time as a constant does not preserve the geodesics in the original 4D black hole spacetime. This calls for a more faithful projection that conserves the Fermat principle of least time.

In this work, we address this question using the Fermat metric, establishing a rigorous analogy between the four-dimensional spacetime metric and a two-dimensional curved surface that faithfully preserves the original four-dimensional lightlike geodesics. From there we design the 2D surface of revolution associated with the non-rotational uncharged Schwarzschild black hole. Because scalar fields in curved spacetime and electromagnetic fields on curved surface share the same equation, namely the Klein-Gordon equation, it is possible to mimic the quasinormal modes of ringing BHs on curved optical microcavities, when their dimensions become comparable to the optical wavelength, far beyond the geometric optics limit explored in earlier works.


In microcavities and microlasers, resonant modes have their backbone along periodic orbits. A famed instance is the whispering gallery modes (WGMs), whose ray counterparts undergo polygonal orbits by total internal reflection along the circular boundary.
Recently, Song et al. constructed non-Euclidean microlaser cavities on a M\"{o}bius strip, and experimentally demonstrated that modes are only sustained on periodic orbits \cite{Song2021}. In these cases, light is confined in the cavity through total internal reflection, with the underlying periodic ray orbit necessarily colliding on the boundary.


Here we show that in non-Euclidean geometry, the spatial curvature may provide a radically different confining mechanism based on an effective ``surface potential" \cite{Costa1981}, which can possibly trap light without involvement of outer boundaries. High-Q bottle micro-resonators based on this concept have been demonstrated on optical glass fiber, with modes localizing on a bulge area \cite{bottle2009}. In this example, however, the effective potential is attractive and the corresponding orbits are stable. This contrasts with the a priori repulsive potential induced by curved spacetime of BHs. 


We develop an analytic approach to identify the quasinormal modes of BHs optical analogue. We demonstrate the existence of a new family of modes, alongside the well-known WGMs, which are confined near the photon sphere. Their spectral and spatial characteristics along the photon ring are unveiled analytically and confirmed through finite difference time domain (FDTD) simulations.
Schwarzschild laser microcavities are fabricated using 3D direct laser writing in dye-doped resin.
Lasing at the PS is demonstrated. Selective pumping is employed to distinguish between PS modes (PSMs) and WGMs and to map their radial profiles, in excellent agreement with the theoretical prediction. This black hole laser paves the way for designing new types of laser microcavities based on non-Euclidean surfaces and potential trapping, inspired by celestial objects.

\section{Theoretical predictions}
\subsection{The photon sphere and its stability}

\begin{figure*}[ptb]
\centering
\includegraphics[width=1\textwidth, trim={0.5cm 0 0 0}, clip]{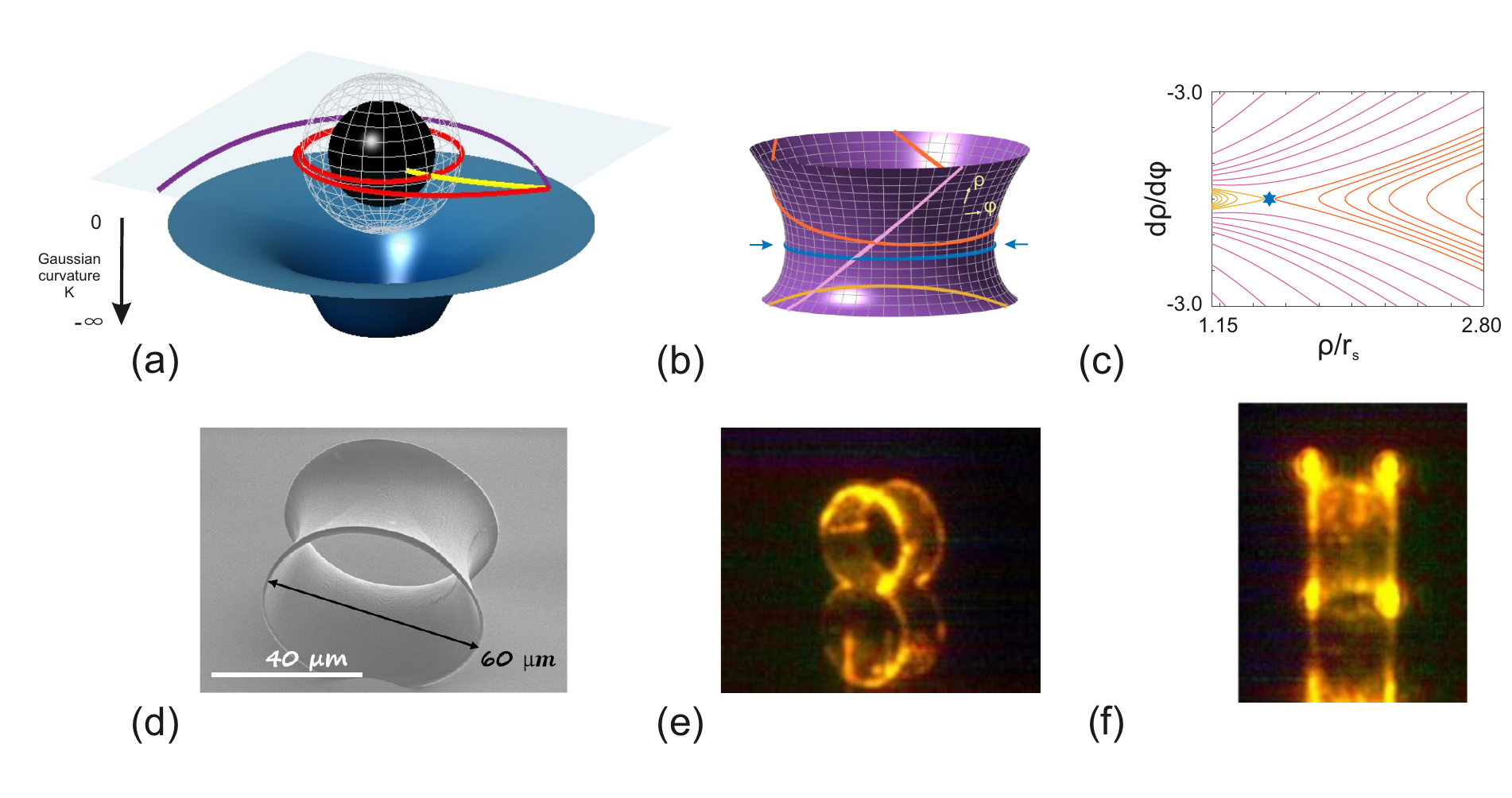}
\caption{(a) Artist view of a black hole (black sphere) and its Gaussian curvature $K$ (bottom surface), which is always negative and becomes nearly zero at very large distances from the black hole. The spherical mesh materializes the position of its photon sphere. The red line represents a light trajectory trapped on the photon sphere, while trajectories that deviate from it are either absorbed (yellow) or deflected and escape (purple). (b) Analog curved surface of a Schwarzschild black hole with reduced dimensionality, featuring coordinates $\rho$ and $\varphi$ along the longitudinal and azimuthal directions, respectively. Various trajectories are shown, with the blue solid line (and blue arrows) representing the photon ring orbit. (c) Poincar\'{e} surface of section of truncated Schwarzschild surface, where lines correspond to trajectories with same color as in (b). The photon sphere trajectory is marked by the blue star. (d) SEM image of the optical microcavity analog fabricated by direct laser writing. (e) and (f) Different optical images of the BH microcavity when uniformly pumped. The microcavity is resting on a glass slide and the recorded image includes both the main object and its reflection.
}%
\label{figure1}%
\end{figure*}

The curved spacetime in the vicinity of a spherically symmetric BH can
be mathematically depicted by the line element $ds^{2}\equiv g_{ij}dx^{i}dx^{j}$ in Schwarzschild coordinates $\left(t, \rho,\theta,\varphi\right)  $ as%
\begin{align}
ds^{2}
&  =-f(\rho)c^{2}dt^{2}+f^{-1}(\rho)d\rho^{2}+\rho^{2}d\theta^{2}+\rho^{2}%
\sin^{2}\theta d\varphi^{2}. \label{Sbh}%
\end{align}
We consider the Schwarzschild metric, with
\begin{equation}
f(\rho)=1-\frac{r_{s}}{\rho}. \label{SchwarzsF}%
\end{equation}
As the simplest spacetime containing a BH, its electric charge, angular momentum and cosmological constant are all vanishing, and its radius $r_{s}$ is determined merely by its mass.

In recent years, significant research has focused on the Fermat principle in general relativity, as it provides a mathematical framework for describing the gravitational lensing effect in astrophysics \cite{Masiello2021, Caponio2011, Perlick2004}. Here, we use it to reduce dimensionality while preserving the geodesic nature of the projected lightlike trajectories. This is performed by applying a conformal transformation of the spatial component of the metrics, $ds^{2}$, into the Fermat metrics, $ds_{\text{Fermat}}^{2}$ \cite{Frankel1979}, defined by
\begin{equation}
ds_{\text{Fermat}}^{2}\equiv\frac{ds^{2}}{\sqrt{-g_{00}}}=f^{-2}(\rho)d\rho^{2}+\rho^{2}f^{-1}(\rho)d\varphi^{2}.
\label{Fermat}%
\end{equation}
The Fermat metric ensures that the Fermat principle is also satisfied in the 3D space without a time coordinate and that geodesics are preserved in the projection process (a detailed derivation can be found in Ref. \cite{Straumann}).
In contrast, the traditional approach of assuming constant time \cite{Bekenstein2014, Ulf2015, Bekenstein2017, Xu2018} fails to reproduce the four-dimensional geodesics.
We further restrict ourselves to the equatorial plane with $\theta=\frac{\pi}{2}$ leveraging on the rotational symmetry.


From there, we construct the corresponding two-dimensional surface of revolution derived from the Schwarzschild metric. Figure \ref{figure1}(b) illustrates the hyperboloid-like representation of the Schwarzschild black hole —hereafter referred to as the ``Schwarzschild surface". (For further details on the construction of this curved surface, see Supplementary Information Section II). The lower boundary is naturally imposed by the event horizon \cite{note}, while an upper boundary is artificially applied; Its location can be chosen arbitrarily; However, it is set parallel to the lower boundary to preserve the rotational symmetry and thereby form an integrable system \cite{note2}. With this method of truncation, the Schwarzschild surface forms a finite non-Euclidean cavity. Microlasers with such a geometry are fabricated by direct laser writing. A scanning electron microscope (SEM) image is shown in Fig. \ref{figure1}(c).

The counterparts of the PS on a curved surface meets the condition obtained from the null or light-like geodesic equations (see Supplementary Information Section I for more details)
\begin{equation}
\frac{\rho}{2}\frac{df(\rho)}{d\rho}-f(\rho)=0. \label{condition}%
\end{equation}
From Eq. (\ref{condition}), one finds that the Schwarzschild surface has one unique circular periodic orbit located exactly at its waist (see blue arrows in Fig.~\ref{figure1}(b)). This periodic orbit corresponds to the well-known photon sphere of the Schwarzschild BH.

In principle, the stability of closed geodesics on a curved surface is quantified by the Jacobi equation associated with Gaussian curvature, which measures the evolution of the distance between two initially close trajectories,
\begin{equation}
\frac{d^{2}\zeta(s)}{ds^{2}}+K(s)\zeta(s)=0.
\end{equation}
This means that the geodesic distance $\zeta(s)$ between two trajectories oscillates when the Gaussian curvature of the surface $K$ is positive, while it diverges when $K$ is negative. In Supplementary Information Section II, we demonstrate that the Gaussian curvature of a Schwarzschild surface is negative everywhere. This is also illustrated in Fig. \ref{figure1}(a) by the blue surface below the BH. Its PS orbit is therefore unstable.
Its instability is also reflected in the Poincar\'{e} surface of section \cite{Genov2009} shown in Fig.~\ref{figure1}(d).

\subsection{The wave equation in the Schwarzschild surface}
In the ringdown stage, the evolution of a scalar perturbation $\Psi$ in a black hole background is described by the wave equation (or massless Klein-Gordon equation \cite{Maggiore2018})
\begin{equation}
\Box\Psi\equiv\frac{1}{\sqrt{g}}\partial_{i}\left(  \sqrt{g}g^{ij}\partial_{j}\Psi\right)
=0, \label{WE}%
\end{equation}
where $\Box$ is the d'Alembert operator, $g=\det\left(  g_{ij}\right)  $, $g^{ij}=\left(  \mathbf{g}^{-1}\right)
_{ij}$ is the element of the inverse matrix of original metric. Interestingly, this equation also describes the propagation of electromagnetic waves constrained on a curved surface embedded in 3D flat spacetime \cite{Batz2008}. The equivalence between these wave equations forms the basis for the analogy between gravitational waves and light waves on curved surfaces. In optics, this can be accomplished by confining light within a thin, curved waveguide, which can be fabricated using direct laser writing technology, as demonstrated later in this article.

\subsection{Existence of Photon Sphere modes}

\begin{figure*}[ptb]
\centering
\includegraphics[width=1\textwidth, trim={0 0 0 0}, clip]{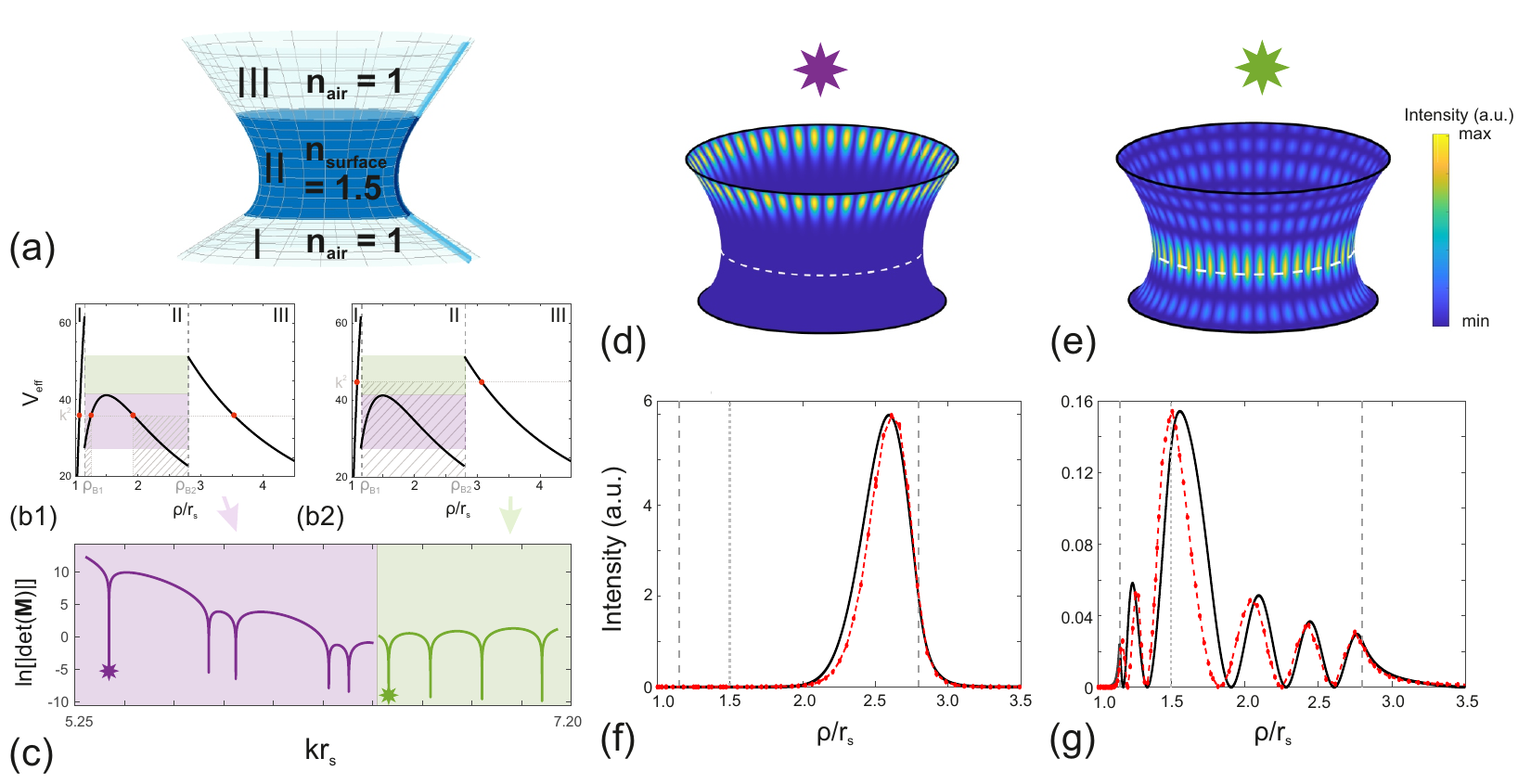}
\caption{(a) Sketch of the truncated Schwarzschild surface (region II, dark blue) and its tangent truncated cones (regions I and III, light blue) at each boundary. (b) Effective potential of light on Schwarzschild surface. Figures (b1) and (b2) correspond to two different values of $k^2$. The purple and green area correspond to the $k^{2}$ range of whispering gallery modes (b1) and photon sphere modes (b2), respectively. The shadow highlights the oscillatory areas of WKB approximation. In (b1), there exist two turning points, denoted by red dots. (c) Log of absolute determinant of the matrix in Eq. (\ref{matrix1}). Quasinormal modes are identified as sharp dips and are divided into two families: whispering gallery modes (WGM, purple, low $kr_s$ values) and photon sphere modes (PSM, green, high $kr_s$ values). The mode with the lowest $kr_s$ value of each family are highlighted by a star symbol in (c) and plotted in (d-g) figures. (d,f) Mode intensity profile of WGM. (e, g) Mode intensity profile of PSM. Their analytical radial intensity profiles (black solid line) highly conform to simulation results (red dashed line) in (e) and (g).}
\label{figure2}%
\end{figure*}

Using this analogy, we will now analytically compute the eigenmodes on Schwarzschild surfaces, which serve as the 2D counterparts of modes from actual black holes.

Thanks to the rotational symmetry of the surface of revolution defined in Eq.~(\ref{Fermat}), we can write the ansatz $\Psi=\rho^{-\frac{1}{2}}\left(1-\frac{r_{s}}{\rho}\right)
^{-\frac{1}{4}}\psi\left(\rho\right)  e^{-il\varphi}e^{ikct}$ to separate variables (see Supplementary Information Section III for details). Here, $l$ is the azimuthal quantum number, which has to be an
integer to fulfill the periodic condition $\Psi(\varphi)=\Psi(\varphi+2\pi)$, $k$
is the to-be-determined eigenwave number, and $\psi\left(\rho\right)  $
satisfies%
\begin{equation}
\frac{d^{2}\psi\left(  \rho\right)  }{d\rho^{2}}+Q_{s}(\rho)\psi\left(
\rho\right)  =0, \label{Schrodinger}%
\end{equation}
where%
\begin{align}
Q_{s}(\rho)  &  =\left(  1-\frac{r_{s}}{\rho}\right)  ^{-2}\nonumber\\
&  \times\left[  k^{2}-\left(  l^{2}+\frac{1}{2}\right)  \left(  \frac{1}%
{\rho^{2}}-\frac{r_{s}}{\rho^{3}}\right)  +\frac{3}{16}\left(  \frac{2}{\rho
}-\frac{r_{s}}{\rho^{2}}\right)  ^{2}\right]  . \label{Qs1}%
\end{align}
Equation (\ref{Schrodinger}) takes a form reminiscent of the Schr\"{o}dinger equation.

The eigenmodes of the open Schwarzschild cavity are solutions of Eq. (\ref{Schrodinger}) in a sandwich structure of air-surface-air. For the purpose of modeling, we assume that the radiation emitted from the boundary of the Schwarzschild cavity is distributed along a truncated cone, whose slope is tangent to its generatrix, as depicted by regions I and III in Fig.~\ref{figure2}(a). This assumption is intuitive, as the rotational symmetry of the Schwarzschild surface ensures that radiation follows its longitudes, which form natural paths in flat space. Additionally, the Gaussian curvature of cones is zero, similar to flat space. A general cone can be described by the metric
\[
ds_{\text{cone}}^{2}=\left(  1+\kappa^{2}\right)  d\varrho^{2}+\varrho
^{2}d\varphi^{2},
\]
where $\varrho$ represents the radius of revolution, and $\kappa$ denotes the slope. The eigenfrequency spectrum of the entire system can be determined by matching the different regions with appropriate interface conditions. In this work, we focus on transverse magnetic modes, where $\Psi$ represents the component of the electric field perpendicular to the surface at each point. Both $\Psi$ and its first derivative are continuous across the I-II and II-III interfaces.




We apply the Wentzel-Kramers-Brillouin (WKB) approximation to solve the Schrödinger-like Eq. (\ref{Schrodinger}) \cite{Bender}. This method yields oscillatory solutions when the spatially varying coefficient $Q_{s}(\rho)>0$ (represented by the hatched areas in Fig. \ref{figure2}(b)) and produces exponentially decaying or diverging solutions when $Q_{s}(\rho)<0$. The points where $Q_{s}(\rho)=0$ are known as turning points, where the WKB approximation breaks down. In such cases, the modified Airy function method can be employed, as it effectively handles the behavior at turning points \cite{MAF}. The sign of $Q_{s}(\rho)$  at any given position $\rho$  is determined by the variable $k$. To facilitate comparison, we define an effective potential from Eq. (\ref{Qs1}) as follows:
\begin{equation}
V_{\text{eff}}(\rho)\equiv\left(  l^{2}+\frac{1}{2}\right)  \left(  \frac{1}%
{\rho^{2}}-\frac{r_{s}}{\rho^{3}}\right)  -\frac{3}{16}\left(  \frac{2}{\rho
}-\frac{r_{s}}{\rho^{2}}\right)  ^{2},
\end{equation}
drawing an analogy with the potential in quantum mechanical systems. As illustrated in Fig.~\ref{figure2}(b), the effective potential $V_{\text{eff}}$ on the Schwarzschild surface resembles a hill, with its peak precisely located at the PS. Conversely, in both tangent cones, $V_{\text{eff}}$ decreases as the distance from the I-II and II-III interfaces increases. It is important to note that our approach follows the ``modes-of-the-universe" framework \cite{Scully1973}, where the field is quantized within the Schwarzschild cavity and its tangent truncated cones. Since energy is conserved throughout the system, the corresponding wavenumbers $k$ are real.

To begin, we search for eigenmodes where
{$k^{2}>V_{\text{eff}}%
^{(II)}(\rho_{\text{PS}})$, which corresponds to the case of Fig.~\ref{figure2} (b2). In this case, we also restrict $k^{2}$ such that $k^{2}<\min\left(  V_{\text{eff}}^{(I)}(\rho_{B1}),V_{\text{eff}%
}^{(III)}(\rho_{B2})\right)$, ensuring the presence of turning points in both Region I and Region III. These turning points divide each region into a decaying area (towards the interface) and an oscillatory area (extending towards infinity). As a consequence, QNMs are more confined between the interfaces, and thus more promising for experimental demonstration.
The wave function in the cavity (region II) can be written as
\begin{equation}
\psi^{(II)}\left(  \rho\right)  =C_{2+}\phi_{2+}(\rho)+C_{2-}\phi_{2-}%
(\rho),\label{kai2}%
\end{equation}
where%
\begin{equation}
\phi_{2\pm}(\rho)=\left[  Q_{s}^{(II)}(\rho)\right]  ^{-\frac{1}{4}}\exp\left(  \pm i
{\displaystyle\int\nolimits_{\rho_{B1}}^{\rho}}
\sqrt{Q_{s}^{(II)}(\rho^{\prime})}d\rho^{\prime}\right)  ,\label{kai3}
\end{equation}
and $C_{2+}$ and $C_{2-}$ are to-be-determined constant coefficients. The stitch of the wave function in these three regions, along with to-be-determined coefficients, can be written in matrix form as (see the Method section)
\begin{equation}
\mathbf{M}\left(
\begin{array}
[c]{c}%
C_{1}\\
C_{2+}\\
C_{2-}\\
C_{3}%
\end{array}
\right)  =0,\label{matrix1}%
\end{equation}%
with $C_{1}$ and $C_{3}$ being the coefficients in Region I and III of the outgoing wave function. To obtain nontrivial solutions, the determinant of the matrix must vanish, i.e., $\det\left(  \mathbf{M}\right)  =0$. As $k$ is the only variable in matrix $\mathbf{M}$, eigen-wave numbers are the zeros of
$\det\left(  \mathbf{M}\right)$, represented as sharp dips in Fig.~\ref{figure2}(c).  Given $l=25$, four solutions are found in the green area of Fig.~\ref{figure2}(c), associated with modes confined by the effective potential within the Schwarzschild cavity. The first mode is shown in Fig.~\ref{figure2}(e) and (g), with maximum intensity near the PS. We identify this mode as the fundamental PS mode. Indeed, the envelope of PS modes in Eq. (\ref{kai3}) conforms to $\left[Q_{s}^{(II)}(\rho)\right]^{-\frac{1}{4}}$, peaking at the PS, while the actual peak of intensity might deviate from the PS due to the oscillatory term. Higher orders of PS modes, with faster oscillations, are reported in Supplementary Information Fig.~S2. 

For quasinormal modes with frequencies down to $k\in\left(  \max\left(
V_{\text{eff}}^{(II)}(\rho_{B1}),V_{\text{eff}}^{(II)}(\rho_{B2})\right) ,V_{\text{eff}}^{(II)}(\rho_{\text{PS}})\right) $, , which corresponds to the case of Fig. \ref{figure2} (b1), two turning points appear in the Schwarzschild cavity, dividing region II into three areas. PS modes can no longer be sustained as the PS falls in the decaying area. Instead, oscillating areas are situated near the two boundaries, resulting in the corresponding quasinormal modes akin to WGMs. A typical WGM solution of Eq. (\ref{kai2}) is shown in Fig.~\ref{figure2}(d) and (f) for comparison. We show in Supplementary Information Fig. S3 the wave functions of all five WGM solutions found in the purple region of Fig.~\ref{figure2}(c). 

\begin{figure*}[ptb]
\centering
\includegraphics[width=1\textwidth, trim={0 0 0 0}, clip]{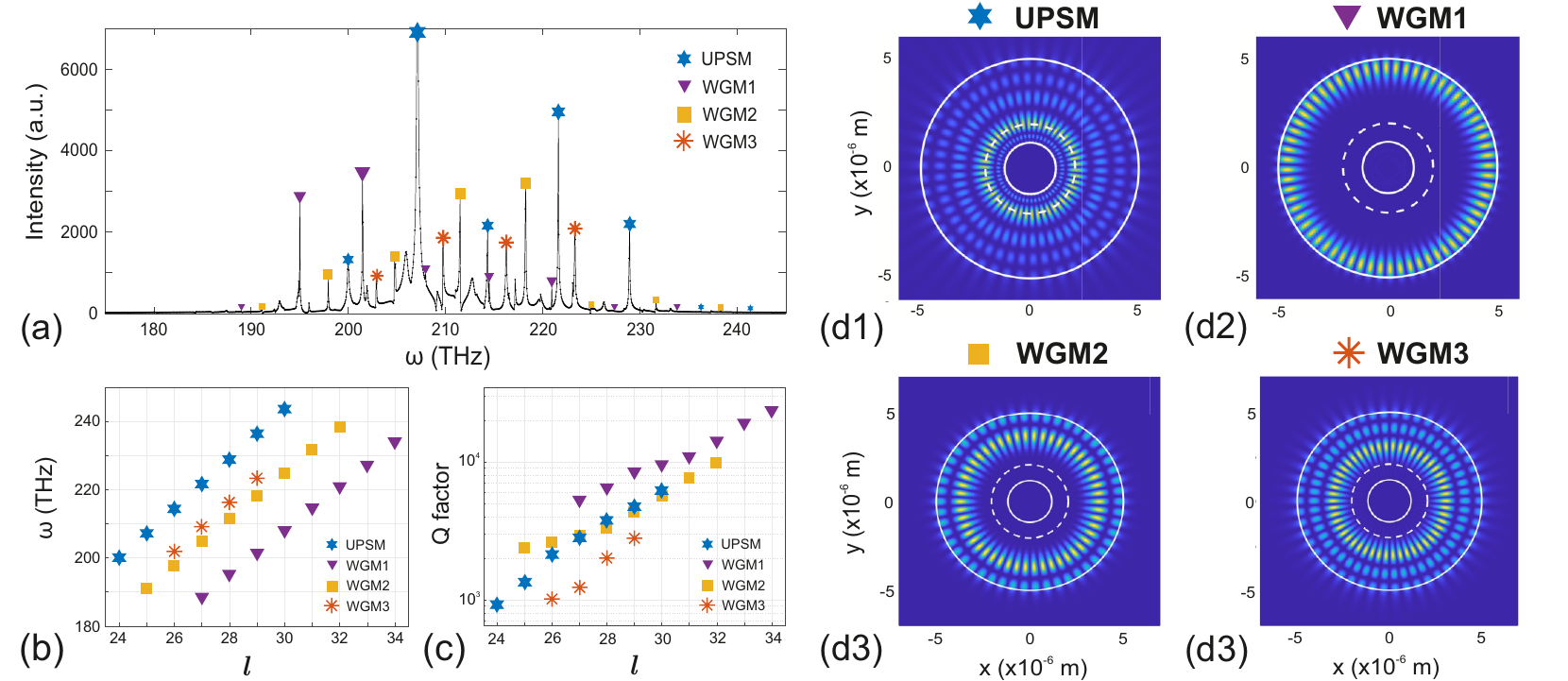}
\caption{(a) Numerical spectrum. Four families of quasinormal modes, including the fundamental unstable photon sphere mode (UPSM) and three whispering gallery modes (WGMs), are identified from (a). (b) Frequencies and (c) Q factors of all the identified modes in (a) with different angular numbers. (d1)-(d4) illustrate the typical intensity distribution of these modes. The PS orbit is indicated by a dashed white line.
}
\label{figure3}%
\end{figure*}

\subsection{Numerical simulations of quasi-normal modes}
To test our theoretical prediction, we perform finite differential time domain (FDTD) simulations and numerically exhibit the eigenmodes in the Schwarzschild cavity. As 3D surfaces could be computationally expensive, here, the simulation is implemented instead in its conformally transformed 2D plane \cite{PNAS}.
Thus the Schwarzschild cavity is represented by a disk with radius 5~{\textmu}m and a radially varying discretized index of refraction.
Perfectly matched layers (PML) surround the simulated area, acting as the open boundary at the inteface II-III. A TM-polarized pulse with central frequency $210$ THz (corresponding to wavelength 1.4~{\textmu}m) and pulse length $30$ fs is applied, and the temporal evolution of the electric field is recorded at different positions. 
The 2D simulations are performed by the Ansys Lumerical FDTD software.

By Fourier transforming the sum of all collected time signals, we obtain the power spectrum in Fig.~\ref{figure3}(a). 
Spectral peaks correspond to resonances of the structure, of which the intensity distribution can be computed. Intensity distribution of four typical quasinormal modes are shown in Fig.~\ref{figure3}(b). Different families of modes are identified from their intensity distribution, corresponding to PS modes when peaked near the waist [Fig.~\ref{figure3}(b1)], and WGMs when confined near the system boundaries, [Fig.~\ref{figure3}(b2)]. The radial intensity distribution can be extracted and compared with theoretical calculations, as illustrated in Fig.~\ref{figure2}(f) and (g). The simulated intensity profiles, represented by red dashed lines, show a high degree of consistency with the analytical results, providing strong validation for the theoretical predictions.

The resonance frequencies of all modes \cite{WGMinside} in Fig.~\ref{figure3}(a) are plotted in Fig.~\ref{figure3}(c) as a function of their azimutal number, $l$, found from the intensity spatial distribution of each of these modes. As expected, different families of modes, including higher order WGMs, follow a linear dependence with $l$.

Finally, we are able to extract the quality factor, $Q=\omega_{0}/\Delta\omega$, of each mode by fitting each spectral peak in Fig.~\ref{figure3}(a) with a Lorentzian function \cite{Qfactor,PML}. The Q factors for different families of mode are shown in Fig.~\ref{figure3}(d) as a function of their azimutal number, $l$. The Q factors of the PS mode family are lower than those of the first-order WGM. This difference arises because WGMs are strongly confined near the outer boundary by total internal reflection, whereas the potential-induced confinement of PSMs does not fully suppress their radiation loss from the truncated structure, as shown in Fig.~\ref{figure3}(b1). 
Nevertheless, the Q factors of PS modes are comparable to or even exceed those of higher-order WGMs (Fig.~\ref{figure3}(d)), highlighting the effectiveness of curvature-induced light confinement in the propagation layer.

\begin{figure*}[hptb]
\centering
\includegraphics[width=1.2\textwidth, trim={5.2cm 0 0.1cm 0}, clip]{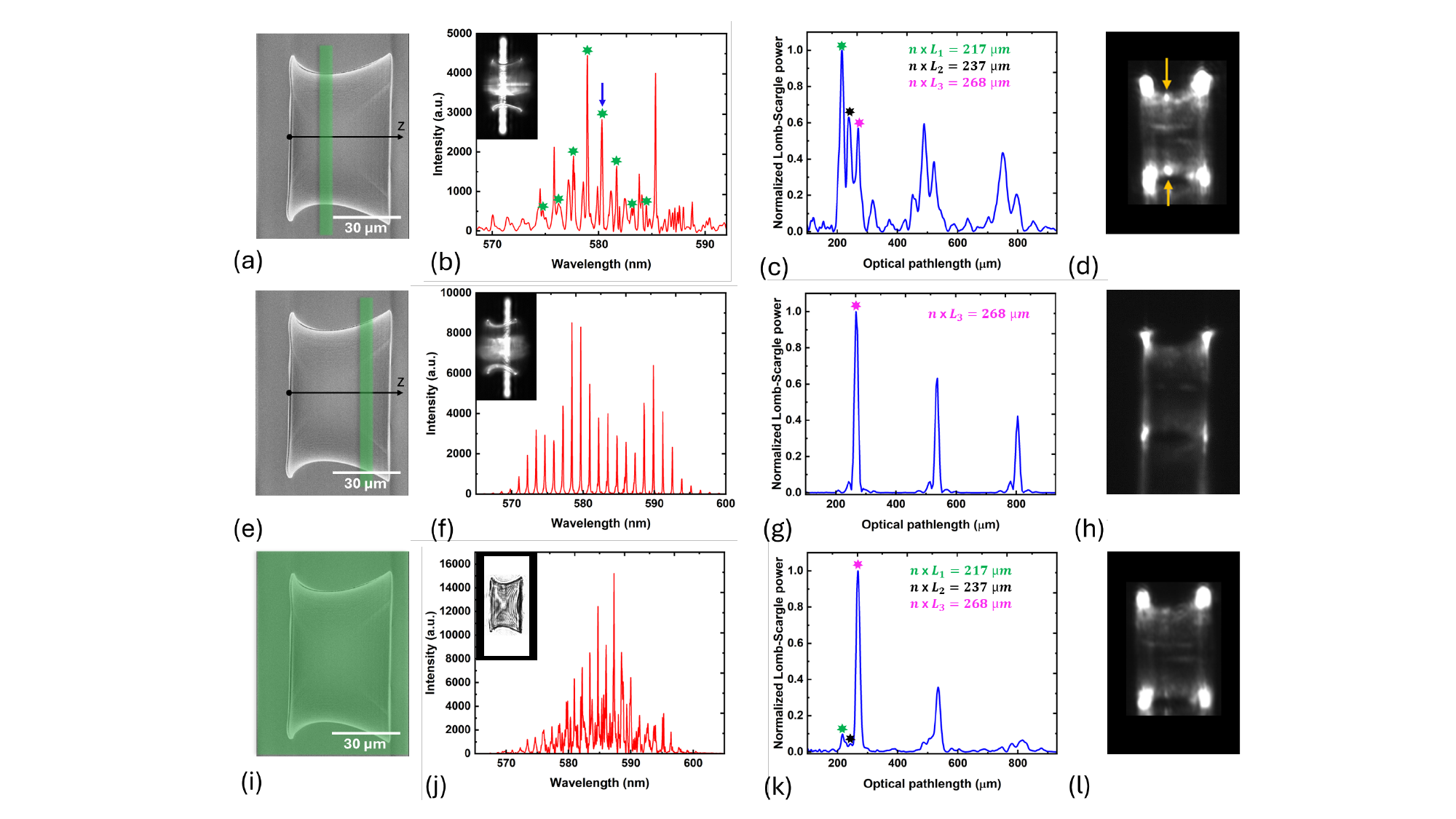}
\caption{
(a) SEM images of the 3D Schwarzschild black hole microcavity laser resting on a glass plate. The z-axis represents the axis of the surface of revolution. The 4 $\mu$m-wide and 95 $\mu$m-long pump stripe, illustrated by the green rectangle, is placed at the waist to selectively excite the photon sphere mode. (b) Emission spectrum of the microstructure when locally pumped near the waist. PS modes are indicated by stars. Inset: top view of the structure with the pump stripe at the waist. (c) Lomb-Scargle periodogram of the emission spectrum in (b). Three peaks are identified in the first group at $nL_1$ = 217 $\mu$m, $nL_2$ = 237 $\mu$m, and $nL_3$ = 268 $\mu$m. (d) Direct observation of the black hole laser emission (the pump wavelength is blocked by a notch filter). Yellow arrows point to the PS lasing mode at the waist. (e) Selective excitation of the WGM confined near the sample edge. (f) Emission spectrum of the laser microcavity when selectively pumped near the WGM location close to the sample edge. Inset: the pump stripe's position. (g) Lomb-Scargle periodogram of the emission spectrum in (f). Only the peak at $nL_3$ = 268 remains. (h) Direct observation of the WGM laser emission. Emission at the waist has disappeared. (i) Uniform pumping of the microcavity laser with a 50 $\mu$m-wide and 95 $\mu$m-long rectangular pump profile. (j) Emission spectrum of the microstructure when uniformly pumped. The inset shows the top view of the structure. (k) Lomb-Scargle periodogram of the emission spectrum in (j), identifying a dominant peak at $nL_1$ = 217 $\mu$m along with a feeble signature of the PS mode and the WGM confined near the left sample edge. (l) Direct observation of lasing microcavity under uniform pumping.
}
    \label{Experiment}
\end{figure*}

\section{Experimental demonstration}
We now proceed to experimentally validate our theoretical predictions on the existence of modes at the PS. The Schwarzschild surface is fabricated using advanced 3D direct laser writing, utilizing Nanoscribe's two-photon polymerization technology. Given the relatively short-lived nature of the predicted optical quasinormal modes, we propose investigating the corresponding lasing modes in a dye-doped organic structure.
This approach would help mitigate the losses and leverage the gain to selectively amplify the modes of the passive system. A SEM image of the fabricated 3D microcavity laser is shown in Fig.~\ref{figure1}(d). The fabrication process, the sample details and the experimental setup are described in the Method section.

\begin{figure*}[t]
\centering
\includegraphics[width=0.93\textwidth]{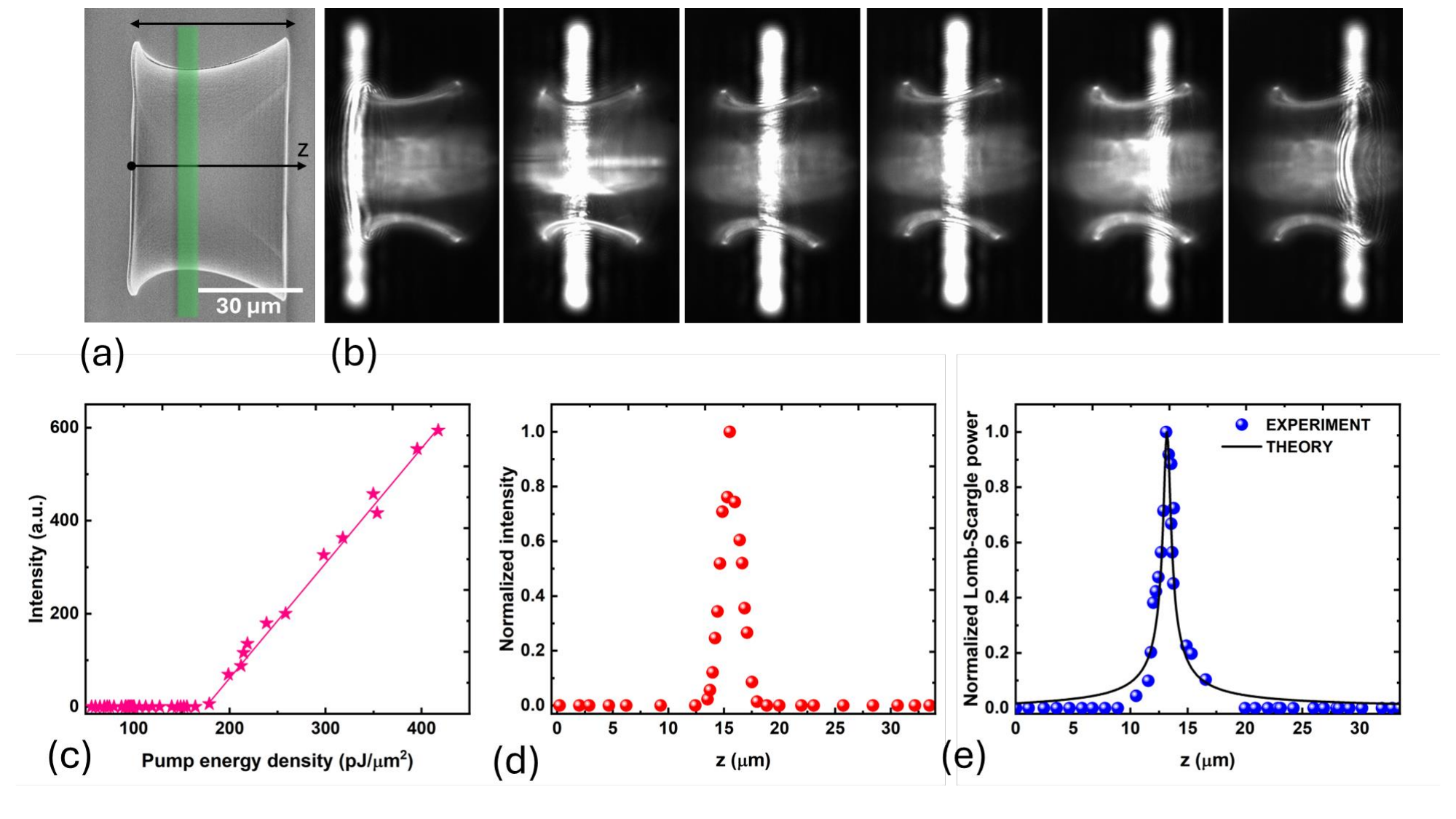}
\caption{(a) SEM image of a Schwarzschild microlaser. The rectangular pump stripe is illustrated with the green rectangle.
(b) Scanning of the pump stripe along the axis of the surface of revolution (z-axis). Few instances of the scanning process.
(c) Laser characteristic of the PS lasing mode at $580.3$ nm under selective pumping.
(d) Spatial intensity distribution of the PS mode, derived from the emission spectrum at 580.3 nm (indicated by the arrow in Fig.~\ref{Experiment}(b)), as the pump stripe is scanned along the z-axis.
(e) Amplitude of the peak at $nL_1$ = 217 $\mu$m (corresponding to PS modes) in the Lomb-Scargle periodogram of Fig.~\ref{Experiment}(c) as the pump stripe is scanned along the z-axis (blue symbols).
The experimental data are compared to the theoretical prediction of the spatial intensity profile of the PS modes with azimuthal number $l=377$ (black line).
}
    \label{scan}
\end{figure*}

\subsection{Experimental evidence of Photon Sphere modes}

With their high Q-factor, WGMs are anticipated to dominate the emission spectrum, as they have a lower lasing threshold than PS modes. To enhance the likelihood of observing PS lasing modes, we propose to pump the microcavity locally, near the PS, away from the edges where WGMs are predominantly confined. Selective pumping has been proposed earlier to selectively excite lasing modes and achieve single mode lasing in random lasers \cite{Bachelard14}. The pump intensity profile of a frequency-doubled Q-switched Nd:YAG laser at 532 nm is shaped by reflection on a spatial light modulator into a 4 $\mu$m-wide and 95 $\mu$m-long narrow stripe positioned at the waist, as shown in  Fig.~\ref{Experiment}(a). The top view of the structure with the pump stripe aligned along the waist is shown in the inset of Fig.~\ref{Experiment}(b). At sufficiently high pump energy, multimode laser emission is observed. The emission spectrum is recorded and shown in Fig.~\ref{Experiment}(b). To identify  periodic features in this spectrum, we applied the Lomb-Scargle algorithm commonly used to characterize periodicity in unevenly-sampled series. The resulting Fourier-like power spectrum is shown in  Fig.~\ref{Experiment}(c). Each peak in the first group corresponds to a family of spectrally regularly-spaced modes confined on a particular orbit with optical path length $nL$. Here, $n = 1.56$ is the group refractive index, obtained from an independent measurement using a reference cuboid microlaser (see Fig.~S4 in the Supplementary Information). The other groups of peaks represent higher harmonics, $mnL$, where $m$ is an integer. Three distinct peaks are seen in the first group, indicating the presence of lasing modes associated with three distinct orbits. The dominant peak corresponds to an optical path length $nL_1$ = 217 $\mu$m. Assuming the orbits are circular, we find a corresponding diameter $D_1$ = 44.3 $\mu$m, which closely matches the waist diameter, $D_W$ = 44.5 $\pm 1$  $\mu$m, providing the first experimental evidence that a family of modes exists at the photon ring of the optical analogue of Schwarzschild black hole.
The corresponding lasing peaks are identified in the emission spectrum of Fig.~\ref{Experiment}(b) and are indicated by stars in the figure. They form a comb of equally-spaced lasing frequencies associated with modes with increasing azimutal number, $l$, as predicted by the theory (Fig.~\ref{figure3}(c)).
PS lasing is directly observed under the microscope, as shown in Fig.~\ref{Experiment}(d), where two lasing spots are seen precisely at the location of the PS.

The two additional peaks seen in the Lomb-Scargle periodogram at $nL_2 = 237\:\mu$m and $nL_3 = 268\:\mu$m are identified as modes on circular orbits with $D_2 = 48.4\:\mu$m and $D_3 = 54.7\:\mu$m. These modes are confined near the sample edges, at $D_L$ =48.9 $\pm 1$  $\mu$m and $D_R$ = 60.0 $\pm 1$ $\mu$m, and are believed to be WGMs. This is confirmed by moving the pump strip to the positions of these modes. For instance, Fig.~\ref{Experiment}(f) and \ref{Experiment}(g) respectively shows the emission spectrum and the periodogram when pumping near the position of orbit $L_3$. The position of the pump stripe on the structure is shown in Fig. \ref{Experiment}(e) and in the inset of Fig. \ref{Experiment}(f).  Only the third peak at $nL_3 = 268\:\mu$m remains in the periodogram, characterizing the periodic structure of the emission spectrum shown in Fig.~\ref{Experiment}(f). As anticipated, the lasing spots at the PS have vanished as can be seen in Fig.~\ref{Experiment}(h). When the pump is moved to other positions, additional WGMs are observed near the sample edges (not shown), corresponding to higher-order WGMs, as predicted by theory (see Fig.~S2 in Supplementary Information).

In contrast to selective excitation, uniform pumping of the microcavity laser would be dominated by WGMs. The emission spectrum with a 50 $\mu$m-wide and 95 $\mu$m-long rectangular pump profile (Fig.~\ref{Experiment}(i)) is presented in Fig.~\ref{Experiment}(j). The corresponding Lomb-Scargle periodogram (Fig.~\ref{Experiment}(k)) reveals a dominant peak at 268 $\mu$m corresponding to the previously identified WGM. A feeble signature of the PS mode and the WGM confined near the sample edge at $D_L$ =48.9 $\pm 1$ is also present. An image of the lasing microcavity under uniform pumping is shown in Fig.~\ref{Experiment}(l) (see also Fig.~\ref{figure1}(e) and Fig.~\ref{figure1}(f)).

\subsection{Photon Sphere mode profile measurement}
We now focus on a specific PS lasing mode that peaks at $580.3$ nm (indicated by the arrow in Fig.~\ref{Experiment}(b)). The peak intensity as a function of pump energy is shown in Fig.~\ref{scan}(c), indicating a laser threshold of 180 $pJ$/$\mu$$m^2$ for this particular mode. To characterize the spatial distribution of this PS mode along the sample axis (z-axis), we record the peak intensity at $580.3$ nm as we slide the pump stripe across the sample in steps of $220$ nm. Few instances of the scanning process are shown in Fig.~\ref{scan}(b). After applying Lucy-Richardson deconvolution, we obtain the spatial distribution of this particular mode shown in Fig.~\ref{scan}(d). The strong confinement observed around the waist position confirms unequivocally that the trapping mechanism is driven by the curvature-induced potential rather than by reflection on the sample edges.

The same scanning process is repeated, but this time, the peak amplitude at $nL_1$ = 217 $\mu$m in the periodogram of Fig.~\ref{Experiment}(c) corresponding to the family of PS modes is plotted against the pump stripe position (Fig.~\ref{scan}(e)). This effectively performs a weighted spatial averaging over the intensity profile of all PS modes. Remarkably, this measurement aligns closely with our theoretical predictions when PS modes are calculated for the actual sample dimensions and corresponding azimuthal number. The azimuthal number $l$ in the experiment can be estimated using the relation $\pi n  D_W = l\lambda$, where $D_W$ is the waist diameter. We find that $l$ ranges from 371 to 383 for all lasing modes identified in the emission spectrum (Fig.~\ref{Experiment}(b)). Considering the average value, $l = 377$, we then calculate analytically the corresponding fundamental PS mode and compare its spatial profile to the measured one (black line in Fig.~\ref{scan}(e)): Excellent agreement in position and spatial extension is obtained between experiment and theory.

\section{Conclusion}
To summarize, we have proposed a dimensionality-reduced optical analog of a 4D Schwarzschild  black hole, which genuinely preserves its light-like geodesics and shares with it the same governing equations. Our analytical investigation, supported by numerical simulations, predicts the existence of modes at the PS of the optical Schwarzschild surface, and we compute their spectral and spatial characteristics. These QNMs, responsible for gravitational radiations in the universe, are visualized here for the first time in 3D-printed Schwarzschild laser microcavities. Selective pumping allows to discriminate between curvature-induced potential-confined PS modes and cavity-trapped WGMs. These modes are confined due to spatial curvature, rather than cavity boundaries, offering new insights into mode confinement mechanisms.

Since the metric in Eq. (\ref{Sbh}) can take different forms, such as by introducing a non-zero cosmological constant, the quantity and position of PSs can be engineered. This paves the way for designing optical cavities and laser devices with novel functionalities.


We believe that black hole analogs, like the one we introduced, can shed new light on current questions in gravitational physics. In particular, perturbations near the event horizon offer a way to test Einstein's theory of general relativity. Here, we present the spatial distribution of the wavefunction, which can be crucial for investigating mode stability \cite{Jaramillo2021,Jaramillo2022}. The stability of the obtained quasinormal mode  spectrum can be further examined using methods such as the pseudospectrum approach \cite{Jaramillo2021}, which is relevant to gravitational-wave analysis \cite{Berti2021} and could help in detecting fundamental modes and overtones \cite{Cotesta2022,spectrum2023}. The analog model discussed in this work may facilitate the study of a wide range of celestial events and phenomena, including black hole spectroscopy \cite{spectroscopy}, Lyapunov exponents \cite{Cardoso2009}, wave dynamics, Hawking radiation, superradiance, and black hole evaporation \cite{Mukhanov}. These analog systems also provide physical testbeds for exploring ideas like gravitational wave echoes \cite{CardosoPani}, and in the future, could lead to studies of backreaction and nonlinear QNM interactions \cite{Nakano:2007cj,Cheung:2022rbm,Mitman:2022qdl,Gleiser:1995gx,Loutrel:2020wbw}.

\section*{Methods}
\subsection{Derivation of coefficient matrix \bf{M}}
Elements of matrix M are wave functions of each region at the boundaries. In Region I and III, as the turning points are far enough from the boundaries, the field decays to a negligible amplitude before reaching the oscillatory region. Therefore we write the wave functions in the decaying areas as
\begin{equation}
\phi^{(I)}\left(  \rho\right)  =C_{1}\phi_{1-}(\rho),\phi^{(III)}\left(  \rho\right)
=C_{3}\phi_{3-}(\rho),
\label{kai5}
\end{equation}
where $C_{1}$ and $C_{3}$ are to-be-determined coefficients,%
\begin{equation}
\phi_{1-}(\rho)=\left[  Q_{s}^{(I)}(\rho)\right]  ^{-\frac{1}{4}}\exp\left(
{\displaystyle\int\nolimits_{\rho_{B1}}^{\rho}}
\sqrt{-Q_{s}^{(I)}(\rho^{\prime})}d\rho^{\prime}\right)  ,
\end{equation}%
\begin{equation}
\phi_{3-}(\rho)=\left[  Q_{s}^{(III)}(\rho)\right]  ^{-\frac{1}{4}}\exp\left(  -%
{\displaystyle\int\nolimits_{\rho_{B2}}^{\rho}}
\sqrt{-Q_{s}^{(III)}(\rho^{\prime})}d\rho^{\prime}\right)  .
\label{kai6}
\end{equation}
Stitching of wave functions Eqs. (\ref{kai2}), (\ref{kai3}), (\ref{kai5})-(\ref{kai6}) at the boundaries leads to
\begin{equation}
\mathbf{M}=\left(
\begin{array}
[c]{cccc}%
\phi_{1-}(\rho_{B1}) & -\phi_{2+}(\rho_{B1}) & -\phi_{2-}(\rho_{B1}) & 0\\
\frac{d\phi_{1-}}{d\rho}(\rho_{B1}) & -\frac{d\phi_{2+}}{d\rho}(\rho_{B1}) & -\frac
{d\phi_{2-}}{d\rho}(\rho_{B1}) & 0\\
0 & \phi_{2+}(\rho_{B2}) & \phi_{2-}(\rho_{B2}) & -\phi_{3-}(\rho_{B2})\\
0 & \frac{d\phi_{2+}}{d\rho}(\rho_{B2}) & \frac{d\phi_{2-}}{d\rho}(\rho_{B2}) &
-\frac{d\phi_{3-}}{d\rho}(\rho_{B2})
\end{array}
\right)  .\label{matrix}%
\end{equation}

\subsection{Sample fabrication}
The laser microcavity is 3D-printed in dye-doped resin, using two-photon polymerization with commercial Nanoscribe lithography GT system. IP-G is used as the Nanoscribe resist and is doped with Pyrromethene 597 (0.5 $\%$ wt) to incorporate gain. Laser printing begins within the glass substrate to securely anchor the microcavity. A dozen of BH microlasers are fabricated on the same glass slide with very good reproducibility. SEM image of a Schwarzschild BH microlaser is shown in Fig.~\ref{figure1}(d), Fig.~\ref{Experiment}(a,e,i) and Fig.~\ref{scan}(a). The waist of the structure has a diameter $D_W$=44.5 $\pm1$ $\mu$m. The diameters of the left (relative to the z-axis defined in Fig.~\ref{Experiment}(a)) and right circular edges are $D_L$=49 $\pm1$ $\mu$m and $D_R$=60 $\pm1$  $\mu$m, respectively.

\subsection{Experimental setup}
The experimental apparatus includes a reflective spatial light phase modulator (SLM) (HES 6001 from Holoeye, pixel size 8.0 $\mu$m), which serves as a secondary display for the computer and can receive a grayscale image of the desired shape as needed. To selectively pump specific areas of the laser microcavity, a MATLAB-generated rectangular grayscale image with a uniform value of 255 is sent to the SLM. The pump laser (Ekspla PL2230 with $\lambda$ = 532 nm, maximum output energy of 28 mJ, pulse duration of 20 ps, repetition rate of 10 Hz) reflects this grayscale image from the SLM. Furthermore, the SLM rotates the pump laser's polarization by 90 degrees in regions where the grayscale value is non-zero. A polarizer positioned after the SLM is optimized to achieve the corresponding amplitude modulation. The pump polarization is aligned with the cavity axis of rotation (z-axis). The setup also includes two sCMOS cameras, one attached to a zoom lens, the other to a fixed-stage microscope (Zeiss AxioExaminer A1) for imaging the 3D cavity from both the side and the top. The shaped pump beam is directed upward, allowing top imaging under the microscope to ensure precise alignment of the pump strip perpendicular to the laser microcavity's axis and to monitor mode profile as it is moved along the structure. Lasing emission is collected by a microscope objective (20 $\times$ from Thorlabs) connected to a high-resolution imaging spectrometer (iHR550 from Horiba, 2400 mm$^{-1}$ density grating, spectral resolution of 20 pm and Synapse camera) via a multimode fiber.

\subsection{Data analysis}
The Lomb-Scargle routine of Matlab, plomb.m, has been used, which yields a Fourier-like power spectrum, but with a better accuracy in the peak positions and a better signal-to-noise ratio than usual discrete Fourier transform algorithms.

The Lucy-Richardson deconvolution is a widely used algorithm for image deblurring. The routine deconvlucy.m from Matlab has been used here to improve the resolution of the mode profile, beyond the width of the pump stripe.

\section*{Acknowledgments}
{This work was done within the C2N micro nanotechnologies platforms. We appreciate precious discussions with Stefan Bittner in Universit\'{e} de Lorraine, Barbara Dietz in Korea University of Science and Technology, Hugo Girin, Xavier Checoury, Joseph Zyss in Universit\'{e} Paris-Saclay, Guillaume Bossard in Ecole Polytechnique, Mordechai(Moti) Segev in Technion, Azriel Z. Genack in City University of New York, Jiaoqing Wang, Sophia Purow in Bar-Ilan University. We also thank Eitan Tsuk in Bar-Ilan University for instruction on using the simulation toolboxes of black holes. C. X. acknowledges the Excellence Fellowship for international postdoctoral researchers funded by Israel Academy of Sciences and Humanities and the Council for Higher Education of Israel. The Bar Ilan group acknowledge the precious help of its lab manager, Leonid Wolfson. }

\section*{Availability of data and materials}
{The datasets used and/or analysed during the current study are available from the corresponding author on reasonable request.}
\section*{Competing interests}
{The authors declare that they have no competing interests.}
\section*{Funding}
{This work was supported by The Israel Science Foundation grants 1871/15, 2074/15, 2630/20 and 1698/22; the United States–Israel Binational Science Foundation NSF/BSF Grant 2015694 and 2020245, the BSF Grant 2022158.  This work was partly supported by the French
RENATECH network and the General Council of Essonne, France.}
\section*{Authors' contributions}
{P.S. and M.L. conceived the idea, designed the research and supervised the project. C.X. performed all numerical simulations, analytical calculations and theoretical analysis. A.S., M.L. and N.B.K. developed the experimental setup, conducted the experiments and performed data analysis. C.X. and A.S. drafted the initial manuscript. C.L. and D.D. fabricated the structures. L.Z., L.W., O.B. actively participated in the discussions and provided valuable insights. All authors reviewed and approved the final manuscript. }

\appendix

\section{Geodesics on Schwarzschild surfaces}

On an arbitrary curved spacetime with a given metric, null geodesics, which are
the natural paths for light rays, follow the full geodesic equation
\begin{equation}
\frac{d^{2}x^{\sigma}}{d\xi^{2}}+\Gamma_{\mu\nu}^{\sigma}\frac{dx^{\mu}}{d\xi
}\frac{dx^{\nu}}{d\xi}=0. \label{adgg}%
\end{equation}
Here $\xi$ is the affine parameter which could be taken as the line element $s$,
$\Gamma_{\mu\nu}^{\sigma}=\frac{1}{2}g^{\sigma\rho}(g_{\rho\mu,\nu}+g_{\rho
\nu,\mu}-g_{\mu\nu,\rho})$ is the Christoffel connection with $g^{\sigma\rho}$
being elements of the inverse of the metric tensor $\mathbf{g}$ and \textquotedblleft%
$,$\textquotedblright\thinspace\ denoting first-order derivatives, $\sigma
,\mu,\upsilon,\rho=1,2$ and Einstein's summation convention is used. By plugging
the metric of Eq. (2) into Eq. (\ref{adgg}) and performing basic algebra, we obtain
Eqs. (3) and (4),%
\begin{align}
\frac{d^{2}\rho}{ds^{2}}-\frac{1}{f(\rho)}\frac{df(\rho)}{d\rho}\left(
\frac{d\rho}{ds}\right)  ^{2}+\left[  \frac{\rho^{2}}{2}\frac{df(\rho)}{d\rho
}-\rho f(\rho)\right]  \left(  \frac{d\varphi}{ds}\right)  ^{2}  &
=0,\label{rouS}\\
\frac{d^{2}\varphi}{ds^{2}}+2\left[  \frac{1}{\rho}-\frac{1}{2f(\rho)}%
\frac{df(\rho)}{d\rho}\right]  \frac{d\rho}{ds}\frac{d\varphi}{ds}  &  =0.
\label{faiS}%
\end{align}

Thanks to its rotational symmetry, geodesics on surfaces of revolution conserve angular momenta. This conserved quantity, $\varepsilon$,
can be obtained by solving Eq. (\ref{faiS}) as%
\begin{equation}
\rho^{2}f^{-1}(\rho)\frac{d\varphi}{ds}=\varepsilon.\label{solutionS}%
\end{equation}
For any free trajectory (i.e., without external force, collision, etc), the
quantity $\varepsilon$ remains constant once the initial conditions,
including initial position and direction, are given. As will be demonstrated
in the following, this property results in different behaviors of geodesics
near stable and unstable periodic orbits. With Eq. (\ref{solutionS}), Eq.
(\ref{rouS}) can be solved by the method of variation of constants,
\begin{equation}
\frac{d\rho}{ds}=\eta f(\rho)\sqrt{1-\frac{\varepsilon^{2}}{\rho^{2}}f(\rho)}.\label{solutionrou}
\end{equation}
Here $\eta=1$ corresponds to the case with positive initial longitudinal
speed, i.e., $\left[  \frac{d\rho}{ds}\right]  _{\text{initial}}>0$, and
$\eta=-1$ corresponds to $\left[  \frac{d\rho}{ds}\right]  _{\text{initial}%
}<0$. Note that $\eta=1$ and $\eta=-1$ refer to two different geodesics. With
Eqs. (\ref{solutionS}) and (\ref{solutionrou}), one readily has%
\begin{equation}
d\varphi=\eta\frac{\varepsilon}{\rho^{2}\sqrt{1-\frac{\varepsilon^{2}}%
{\rho^{2}}f(\rho)}}d\rho,
\end{equation}
by which trajectories of geodesics are determined.

\section{Gaussian curvature of surfaces of revolution}

\begin{figure}[ptb]
\centering
\includegraphics[width=7cm]{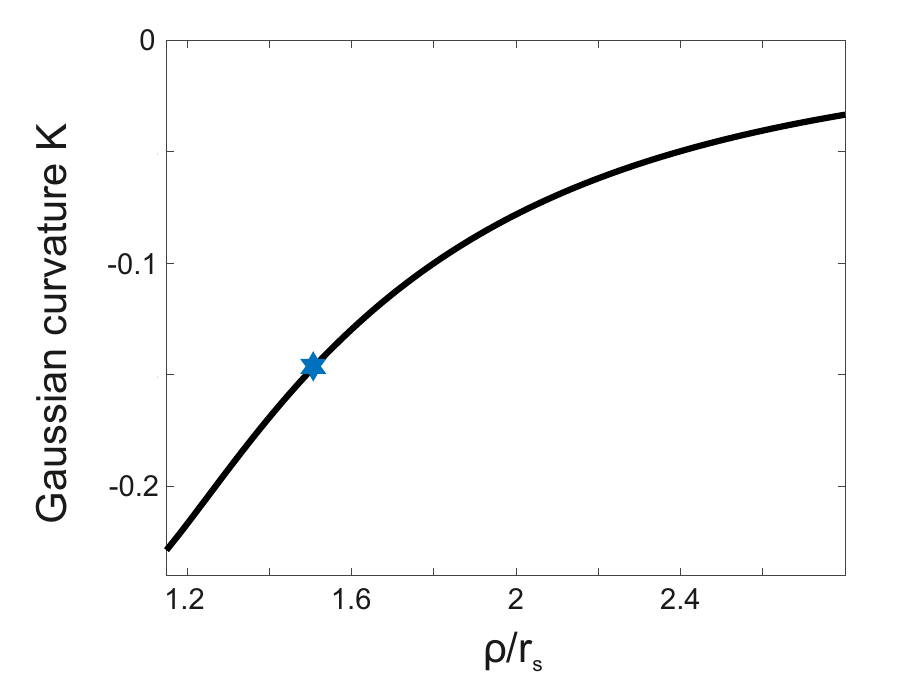}\caption{Gaussian curvature $K$ of a Schwarzschild surface. The blue hexagon denotes the unstable photon sphere.}%
\label{figureS1}%
\end{figure}

Points on an arbitrary surface of revolution can be written as $\mathbf{s}%
=\left[  R(\rho)\cos\varphi,R(\rho)\sin\varphi,H(\rho)\right]  $.
Geometrically, at each point, Gaussian curvature is viewed as product of the
two principal curvatures $\varkappa_{1}$ and $\varkappa_{2}$, where the two
principal curvatures are the reciprocal of principal radii of curvature, which
are the radii of maximal and minimal tangent circles. Mathematically, Gaussian
curvature can be obtained by%
\begin{equation}
K=\frac{eg-f^{2}}{EG-F^{2}},
\end{equation}
where $E,F,G$ are coefficients of the first fundamental forms, and $e$, $f$,
$g$ are coefficients of the second fundamental forms. Specifically, $E$, $F$,
$G$ are defined as
\begin{equation}
E=\frac{\partial\mathbf{s}}{\partial\rho}\cdot\frac{\partial\mathbf{s}}{\partial\rho}=\left[  \frac{dR(\rho)}{d\rho}\right]  ^{2}+\left[
\frac{dH(\rho)}{d\rho}\right]  ^{2},\label{eee}%
\end{equation}%
\begin{equation}
F=\frac{\partial\mathbf{s}}{\partial\rho}\cdot\frac{\partial\mathbf{s}}{\partial\varphi}=0,
\end{equation}%
\begin{equation}
G=\frac{\partial\mathbf{s}}{\partial\varphi}\cdot\frac{\partial\mathbf{s}%
}{\partial\varphi}=R^{2}(\rho),
\label{ggg}%
\end{equation}
the subscript denotes taking the derivative of the vector over the subscripted
variable. $E$, $F$, $G$ are indeed the elements of the metric of this surface
$g_{11}$, $g_{12}(g_{21})$, $g_{22}$, respectively, which implies that given a
metric of a curved surface, one can construct the surface from Eq. (\ref{eee})
and (\ref{ggg}). Furthermore, $e$, $f$, $g$ are computed by%
\begin{align}
e &  =\frac{\partial^{2}\mathbf{s}}{\partial\rho^{2}}\cdot\mathbf{\hat{n}}\\
&  =\frac{1}{\sqrt{E(\rho)}}\left[  -\frac{dH(\rho)}{d\rho}\frac{d^{2}R(\rho
)}{d\rho^{2}}+\frac{dR(\rho)}{d\rho}\frac{d^{2}H(\rho)}{d\rho^{2}}\right]  ,
\end{align}%
\begin{equation}
f=\frac{\partial^{2}\mathbf{s}}{\partial\rho\partial\varphi}\cdot
\mathbf{\hat{n}}=0,
\end{equation}%
\begin{equation}
g=\frac{\partial^{2}\mathbf{s}}{\partial\varphi^{2}}\cdot\mathbf{\hat{n}%
}=\frac{R(\rho)}{\sqrt{E(\rho)}}\frac{dH(\rho)}{d\rho},
\end{equation}
where
\begin{eqnarray}
\mathbf{\hat{n}} &  =&\frac{\frac{\partial\mathbf{s}}{\partial\rho}\times
\frac{\partial\mathbf{s}}{\partial\varphi}}{\left\vert \frac{\partial
\mathbf{s}}{\partial\rho}\times\frac{\partial\mathbf{s}}{\partial\varphi
}\right\vert }\nonumber\\
&  =&
[  -\frac{1}{\sqrt{E(\rho)}}\frac{dH(\rho)}{d\rho}\cos\varphi
,-\frac{1}{\sqrt{E(\rho)}}\frac{dH(\rho)}{d\rho}\sin\varphi,   \nonumber\\
&-&\frac{1}{\sqrt{E(\rho)}}\frac{dH(\rho)}{d\rho} ]
\end{eqnarray}

is the unit normal vector. For simplification, we can use coefficients of the
first fundamental forms to represent $R(\rho)$ and $H(\rho)$, and the Gaussian
curvature is
\begin{equation}
K(\rho)=E^{-2}G^{-1}\left[  E\left(  \frac{1}{4}G^{-1}G^{\prime2}-\frac{1}%
{2}G^{\prime\prime}\right)  +\frac{1}{4}G^{\prime}E^{\prime}\right].\label{K1}
\end{equation}

For the Fermat metric of a Schwarzschild-type curved space, $E(\rho)=f^{-2}%
(\rho),G(\rho)=\rho^{2}f^{-1}(\rho)$. Substituting it into Eq. (\ref{K1}%
), after tedious calculation, we have
\begin{equation}
K(\rho)=-\frac{1}{4}\left[  \frac{df\left(  \rho\right)  }{d\rho}\right]
^{2}+\frac{1}{2}f\left(  \rho\right)  \frac{d^{2}f\left(  \rho\right)  }%
{d\rho^{2}}.
\end{equation}
For a Schwarzschild black hole, $f=1-\frac{r_{s}}{\rho}$, and the Gaussian curvature
is%
\begin{equation}
K_{\text{S}}(\rho)=-\frac{r_{s}}{\rho^{3}}+\frac{3r_{s}^{2}}{4\rho^{4}},
\end{equation}
which can be easily proved negative everywhere.

\begin{figure*}[ptb]
\centering
\includegraphics[width=15cm]{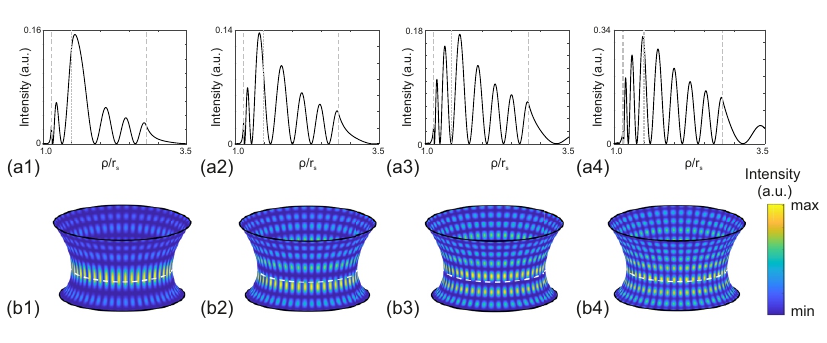}\caption{Different orders of photon sphere modes. (a) Effective potential Schwarzschild surface as shown in the main text, whose range of square of eigen-wave number $k^{2}$ is limited by the orange dot lines. (b) Term $Q(\rho)$ of different regions. (c) Determinant of coefficient matrix, whose zeros correspond to photon sphere modes, denoted by orange stars. Intensity distribution of these four photon sphere modes along $\rho$ direction (d), on Schwarzschild surface (e) and on 2D transformed plane (f). The boundaries of Schwarzschild surface are denoted by gray dashed lines (d) and black (e)/white (f) solid lines, while the photon sphere is denoted by gray dot lines (d) and white dashed lines (e,f).}%
\label{figureS4}%
\end{figure*}

\section{Variable separation of the wave equation}
In this section, we derive the Eqs. (9) and (10) by performing the variable
separation method on the wave equation, which, written in curvilinear coordinates,
is%
\begin{equation}
\Box\Psi\equiv\frac{1}{\sqrt{g}}\partial_{i}\left(  \sqrt{g}g^{ij}\partial_{j}\Psi\right)
=0,\label{WE}%
\end{equation}
where $g^{ij}$ is the element of the inverse matrix of $g_{ij}$, and $g$ is the
determinent of $g_{ij}$. For 2D Schwarzschild surfaces, which are described by the Fermat metric Eq.
(3), as they are embedded in 3D flat spacetime, a flat time term should be
added to obtain a full 3D spacetime metric as%
\begin{equation}
ds^{2}=-c^{2}dt^{2}+f^{-2}(\rho )d\rho ^{2}+\rho ^{2}f^{-1}(\rho )d\varphi
^{2}.  \label{3DDD}
\end{equation}%
Substituting Eq. (\ref{3DDD}) into Eq. (\ref{WE}), we have the wave equation
\begin{equation}
\rho^{-1}f^{\frac{3}{2}}\left(  \rho\right)  \frac{\partial}{\partial\rho
}\left[  \rho f^{\frac{1}{2}}\left(  \rho\right)  \frac{\partial\Psi}%
{\partial\rho}\right]  +\rho^{-2}f\left(  \rho\right)  \frac{\partial^{2}\Psi
}{\partial\varphi^{2}}-\frac{1}{c^{2}}\frac{\partial^{2}\Psi}{\partial t^{2}%
}=0,\label{WE2}%
\end{equation}
Here a time-relevant term $-c^{2}dt^{2}$ is added to the metric Eq. (3), as the surface is embedded in flat spacetime. Assuming variable $t$ can be
separated from other variables, we write wave function $\Psi(t,\rho
,\varphi)=T(t)Y(\rho,\varphi)$. Substituting it into Eq. (\ref{WE2}), we have
\begin{eqnarray}
\frac{f^{\frac{3}{2}}\left(  \rho\right)  }{\rho Y(\rho,\varphi)}
\frac{\partial}{\partial\rho}\left[  \rho f^{\frac{1}{2}}\left(  \rho\right)
\frac{\partial Y(\rho,\varphi)}{\partial\rho}\right]  &+&\nonumber\\
\frac{f\left(
\rho\right)  }{\rho^{2}Y(\rho,\varphi)}\frac{\partial^{2}Y(\rho,\varphi
)}{\partial\varphi^{2}}&=&\frac{1}{T(t)}\frac{1}{c^{2}}\frac{\partial^{2}
T(t)}{\partial t^{2}}.\label{11}
\end{eqnarray}
As Eq. (\ref{11}) works for arbitrary $t,\rho$ and $\varphi$, the only
possibility is that both side equal to a constant, say, $-k^{2}$. Therefore,
one has
\begin{equation}
\frac{\partial^{2}T(t)}{\partial t^{2}}+k^{2}c^{2}T(t)=0\label{tt}%
\end{equation}%
\begin{eqnarray}
\frac{f^{\frac{3}{2}}\left(  \rho\right)  }{\rho}\frac{\partial}{\partial\rho
}\left[  \rho f^{\frac{1}{2}}\left(  \rho\right)  \frac{\partial
Y(\rho,\varphi)}{\partial\rho}\right]  &+&\frac{f\left(  \rho\right)  }{\rho
^{2}}\frac{\partial^{2}Y(\rho,\varphi)}{\partial\varphi^{2}}\nonumber\\
&+&k^{2} Y(r,\varphi)=0.\label{YY}
\end{eqnarray}
Eq. (\ref{tt}) leads to the oscillating wave solution $T(t)=e^{ickt}$. For Eq.
(\ref{YY}), due to the rotational symmetry of surfaces of revolution, variable $\rho$ and $\varphi$ can be separated as
$Y(\rho,\varphi)=R(\rho)\Theta(\varphi)$. Substituting it into Eq. (\ref{YY}),
one has%
\begin{equation}
\frac{\rho f^{\frac{1}{2}}\left(  \rho\right)  }{R(\rho)}\frac{d}{d\rho
}\left[  \rho f^{\frac{1}{2}}\left(  \rho\right)  \frac{dR(\rho)}{d\rho
}\right]  +\frac{\rho^{2}}{f\left(  \rho\right)  }k^{2}=-\frac{1}%
{\Theta(\varphi)}\frac{d^{2}\Theta(\varphi)}{d\varphi^{2}}\equiv l^{2}.
\end{equation}
i.e.,%
\begin{equation}
\frac{d^{2}\Theta(\varphi)}{d\varphi^{2}}+l^{2}\Theta(\varphi)=0,\text{
\ \ }\Theta(\varphi)=e^{il\varphi},
\end{equation}%
\begin{equation}
\rho f^{\frac{1}{2}}\left(  \rho\right)  \frac{d}{d\rho}\left[  \rho
f^{\frac{1}{2}}\left(  \rho\right)  \frac{dR(\rho)}{d\rho}\right]  +\left[
\frac{\rho^{2}}{f\left(  \rho\right)  }k^{2}-l^{2}\right]  R(\rho)=0.
\end{equation}
To get the Schr\"{o}dinger's equation, we apply the ansatz $R(\rho)=\rho^{-\frac{1}{2}%
}f^{-\frac{1}{4}}(\rho)\psi(\rho)$. After tedious algebraic calculation, one
has%
\begin{eqnarray}
\frac{d^{2}\psi(\rho)}{d\rho^{2}}+\nonumber\\
\frac{\rho^{2}}{\left(  \rho-r_{s}\right)
^{2}}\left[  k^{2}-\left(  l^{2}+\frac{1}{2}\right)  \frac{\rho-r_{s}}%
{\rho^{3}}+\frac{3}{16}\frac{\left(  2\rho-r_{s}\right)  ^{2}}{\rho^{4}%
}\right]  \psi\left(  \rho\right)  =0\nonumber\\
\end{eqnarray}
which corresponds to Eqs. (9) and (10) in the main text.

\begin{figure*}[ptb]
\centering
\includegraphics[width=15cm]{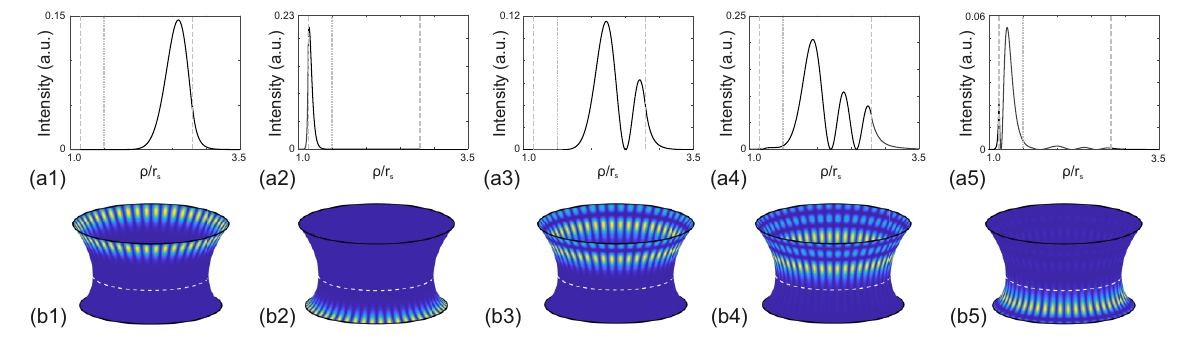}\caption{Five whispering gallery modes of a Schwarzschild cavity. The eigenfrequency spectrum of whispering gallery modes is shown in Fig. 2(c) in the main text.}%
\label{figureS2}%
\end{figure*}

\begin{figure*}[ptb]
\centering
\includegraphics[width=16cm]{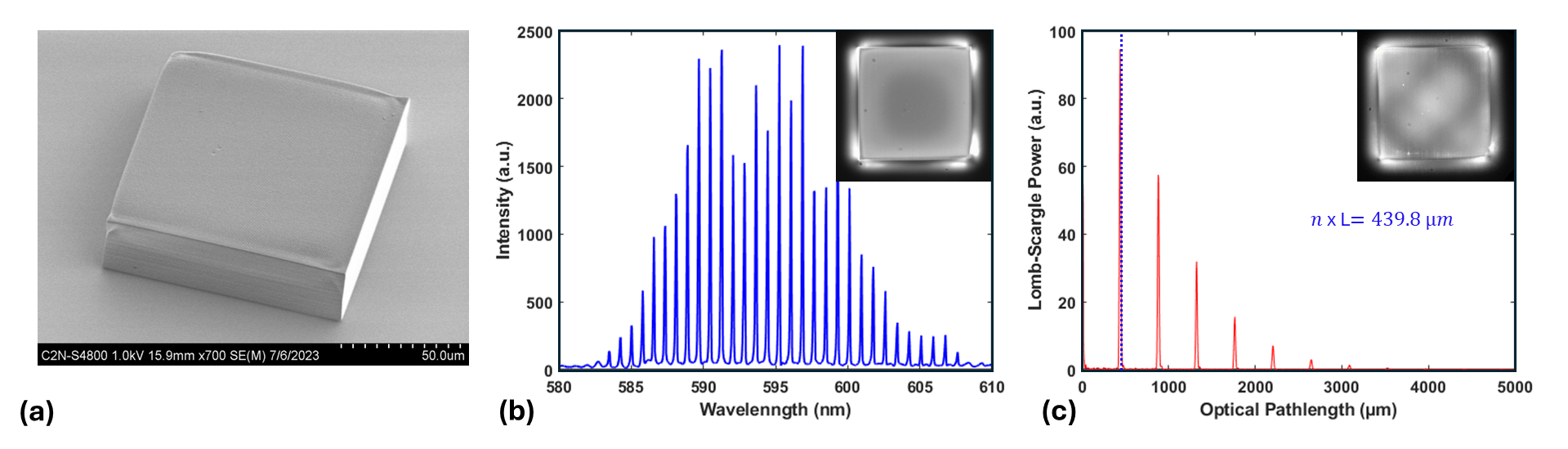}\caption{(a) SEM image of a cuboid microlaser. (b) Emission spectrum of a cuboid microlaser when pumped with a square-shaped pump of dimension 120 $\mu$m x 120 $\mu$m. Inset of the figure shows the top view of cuboid microlaser under white light illumination, with focus on the upper surface. (c) Fourier transform of the emission spectrum in (b). Inset of the figure shows the top view of the cuboid microlaser when pumped with laser.}%
\label{suppl1}%
\end{figure*}

\section{Measurement of the refractive index}

To obtain the group refractive index of the 3D laser microcavities, we 3D-print a cuboid microlaser of side length $a$=100 $\mu$m in the same dye-doped resin. A SEM image of the fabricated structure is shown in Fig. \ref{suppl1}(a). Top view of the cuboid laser under white light, with focus on the upper surface is shown in the inset of Fig. \ref{suppl1}(b). The cuboid microlaser is then pumped with a square-shaped pump profile of 120 $\mu$m x 120 $\mu$m dimension. When the pump power is sufficiently high, we observe an emission spectrum with regularly spaced peaks as shown in Fig. \ref{suppl1}(b). The comb-like spectrum corresponds to a diamond periodic orbit excited in the cuboid laser \cite{OE-3D}. The inset of Fig. \ref{suppl1}(c) shows the top view of the cuboid laser when pumped with  square-shaped pump. Fig. \ref{suppl1}(c) shows the Fourier transform of the emission spectrum, which consists of regularly spaced harmonics. Its first peak at $nL$=440 $\mu$m gives the optical path length corresponding to the diamond orbit, where $L=2\sqrt{2}a$ is the geometrical length of the orbit and $n$ is the group refractive index of the cuboid laser.
For the given optical path length, the group refractive index is calculated as $n$=1.56. Appendix V of Ref. \cite{Song2021} discuss the precision of this measurement.


\begin{thebibliography}{99}
\bibitem{note} Other shapes of the upper boundary can be further explored. For example, a non-parallel upper boundary can mitigate the lasing of WGMs.
\bibitem{WGMinside}
In practice two other WGM families are identified numerically, which are located at the inner boundary. Because of absence of perfectly matched layer in the center, Q factors of these modes are not accurate. Thus these modes are not shown. For more details, see Supp. Info. Fig. S2.
\bibitem {PML}
Indeed, absorption from PML results in the non-vanishing width $\Delta\omega$ of spectral peaks. This information was not accessible through the theoretical method, as details about the modes of the truncated Schwarzschild surface could not be extracted solely from the modes of the universe.
\end{thebibliography}
\end{document}